\documentstyle[aps,epsf]{revtex}
\def\bea{\begin{eqnarray}}
\def\eea{\end{eqnarray}}
\def\nn{\nonumber}
\def\h{\hspace*{.5mm}}
\def\5{\hspace*{5mm}}

\def\bk{{\bf k}}
\def\dr{{\rm d}{\bf r}}
\def\dk{{\rm d}{\bf k}}
\def\p{\partial}
\def\d{\delta}

\def\s{\sigma}
\def\la{\langle}
\def\ra{\rangle}
\begin{document}
\title{Magnetic Instability in a Parity Invariant 2D Fermion System}
\author{M.\h\/Eliashvili and G.\h\/Tsitsishvili}
\address{Department of Theoretical Physics\\
A.\h\/Razmadze Institute of Mathematics\\
Aleksidze 1, Tbilisi 380093 Georgia\\
{\rm e-mail: simi@rmi.acnet.ge}}
\maketitle
\begin{abstract}
\begin{center}
\parbox{14cm}{
We consider the parity invariant QED$_{2+1}$ where the matter is
represented as a mixture of fermions with opposite spins. It is argued that
the perturbative ground state of the system is unstable with respect to the
formation of magnetized ground state. Carrying out the finite temperature
analysis we show that the magnetic instability disappears in the high
temperature regime.}
\end{center}
\end{abstract}

\section{Introduction}

Interest to the (2+1)-dimensional field theory is heated up by its successful
applications to the problems of planar condensed matter systems like the
quantum Hall fluid and unconventional superconductors. One of the relevant
issues concerns the phenomenon of spontaneous magnetization in the systems of
planar fermion matter.

Different authors\cite{1,2,3,4,5,6} have given convincing arguments that the
perturbative ground state (the one with $F_{\mu\nu}=0$) of a 2D dense fermion
matter exhibits the magnetic instability. In other terms there does exist the
true ground state with the nonzero value of a proper magnetic field.

The peculiar property of planar physics is that spin is not quantized, and
spin-up and spin-down states belong to the different representations of the
rotation group $SO(2)$. Remark, that the existence of magnetized ground state
was demonstrated in the systems of fermions with one and the same spin
orientation. Hence one can propose the simple mechanism explaining this
phenomenon: matter consisting entirely of either up or down spin particles
possesses nonvanishing spin density leading to a proper magnetization of
the system.

In the present paper we study the double-spin model (also referred to as
duplicated) where the matter represents the mixture of opposite spin fermions.
Such a consideration can be motivated by the observation$^7$ that the magnetic
properties of duplicated and single-spin (up or down) systems can be quite
different.

In the rest part of this section we remind some principal aspects of
spontaneous magnetization in the single-spin system. In Section 2 we perform
the perturbative study of duplicated system and find out magnetic instability
of perturbative ground state. The mechanism of instability in the double-spin
model turns out to be different from that in the single-spin systems.
Some necessary calculations are placed in the Appendix.

The basic object of our account is the finite temperature effective
action $W[A]$ which arises after integrating out the fermion fields.
In the Matsubara formalism it is given by
\bea
\exp\left\{-W[A]\right\}\h=\int D\psi^*D\psi
\h\exp\left\{-\int L\h d\tau\dr\right\},\nn
\eea
where $L$ is the Euclidean Lagrangian, $0<\tau<T^{-1}$, and
$T$ is the temperature.

Consider the nonrelativistic fermion matter interacting with $U(1)$ gauge
field. The corresponding Euclidean Lagrangian is given by
\bea
L\h=\h\h\frac14\h\h F_{\mu\nu}F_{\mu\nu}-\h
ie\rho A_\tau\h+\h\psi^*(\p_\tau+\h ieA_\tau)\psi\h+\nn
\eea
\bea
+\h\h\frac{1}{2m}\h\h|(\p_k+ieA_k)\psi|^2\h
-\h\h\frac{\s e}{2m}\h\h B\h\psi^*\psi\h-\h\mu\h\psi^*\psi\h,\nn
\eea
where $\rho$ is the neutralizing charge density, $\s$ describes the spin
degree of freedom, magnetic field is defined as $B=\p_2A_1-\p_1A_2$, and
$\mu$ is the chemical potential.

The linear part of $W[A]$ appears as
\bea
W_1\h=\h e\int\{i\h[\la\psi^*\psi\ra\h-\h\rho]
\h A_\tau\h-\h\h\frac{\s}{2m}\h\h\la\psi^*\psi\ra\h B
\}d\tau\dr\h,\nn
\eea
where the symbol $\la\cdots\ra$ denotes the thermal
average corresponding to $F_{\mu\nu}=0$.

From the last expression one can extract the principal conclusion concerning
the magnetic properties of the single-spin system. Due to the term, linear in
the magnetic field, the extremum of $W[A]$ will be realized for some $B\ne0$.
The common conclusion achieved in the finite temperature analysis\cite{5,6}
of single-spin models, implies that the spontaneous magnetization persists
at any temperatures. The reason is quite simple: the finite spin density
$(\s/2)\la\psi^*\psi\ra$ creates the proper magnetization of the system
\bea
M\h=\h-\h\h\frac{\d W_1}{\d B}\h\h=\h
\h\frac{\s e}{2m}\h\h\la\psi^*\psi\ra\h,\nn
\eea
which survives at any temperatures, provided the particle density is finite.

Concerning the $A_\tau$-term, it leads to the electric field corresponding
to a local charge distribution $e\la\psi^*\psi\ra-e\rho$.

\section{Double-Spin System}

In this section we consider the duplicated model where the matter represents
the mixture of opposite spin ($\pm\s$) fermions.

The system is assumed to be in contact with a particle reservoir which keeps
the total particle density fixed and guarantees the chemical equilibrium
between spin-up and spin-down subsystems. This can be realized by equalizing
the corresponding chemical potentials.
Then $\la\psi_\uparrow^*\psi_\uparrow^{}\ra
=\la\psi_\downarrow^*\psi_\downarrow^{}\ra$, and the linear part of $W$
appears as
\bea
W_1\h=\h i\h e\int[\h\la\psi_\uparrow^*\psi_\uparrow^{}
\ra\h+\h\la\psi_\downarrow^*\psi_\downarrow^{}\h\ra\h
-\h\rho\h]\h A_\tau\h d\tau\dr\h,\nn
\eea
where the term linear in the magnetic field is absent since the duplicated
model is parity invariant.\footnote{Magnetic field defined in 2+1
dimensions is a pseudoscalar. Therefore, the spin interaction leads to a
parity violation in single-spin systems. Provided spin-up and spin-down
fermions are interchanged under the parity transformation the duplicated
model is parity invariant.} Therefore, the configuration with $B=0$ represents
the extremum of the corresponding effective action. The prior question which
arises in this connection is the one concerning the stability of this extremum.
This issue can be studied in terms of the two-point functions calculated within
the Gaussian approximation. Considering these functions as integral operators
we can introduce the corresponding eigenvalues which characterize the spectrum
of gauge field fluctuations. Stability of the perturbative ground state
requires the non-negative definiteness of this eigenvalue spectrum. If the
spectrum contains at least one negative eigenvalue, then one is faced with an
instability leading to the formation of the nontrivial ground state. In what
follows we adopt this criterion and search for the negative eigenvalues of the
two-point functions determining the Gaussian part. The later is given by
\bea
W_2\h=\h\h\frac14\int F_{\mu\nu}F_{\mu\nu}\h dx\h+\h\h\frac{e^2m}{2}
\int A_\mu(x_1)G_{\mu\nu}(x_1-x_2)A_\nu(x_2)\h dx_1dx_2\h,\nn
\eea
where $dx=d\tau\dr$, and $G_{\mu\nu}(x)$ are the current correlation
functions. In the Fourier representation they look as
\bea
G_{\tau\tau}\h=\h-\int\frac{\dk'}{2\pi^2}\h\h
\frac{f(\beta E_+)\h-\h f(\beta E_-)}{(\bk\bk')^2\h+\h m^2\xi^2}\h\h(\bk\bk')\h,\nn
\eea
\bea
G_{\tau i}\h=\h G_{i\tau}\h=\h-\h\h
\frac{\xi\h k_i}{\bk^2}\h\h G_{\tau\tau}\h,\nn
\eea
\bea
G_{ij}\h=\h\d_{ij}\h\h\frac{\xi^2}{\bk^2}\h\h
G_{\tau\tau}\h+\h\h\frac{1}{m^2}\h\h
(\h\bk^2\d_{ij}\h-\h k_ik_j)\h G\h,\nn
\eea
\bea
G\h=\int\frac{\dk'}{2\pi^2}\h\h
\frac{f(\beta E_+)\h-\h f(\beta E_-)}{(\bk\bk')^2\h+\h m^2\xi^2}\h\h(\bk\bk')
\left\{\frac{\bk'^2}{\bk^2}\h-\h2\h\h\frac{(\bk\bk')^2}{\bk^4}
\h\h+\h\h\frac{\s^2}{4}\right\},
\eea
where $\xi=2\pi Tn$ with $n=0,\pm1,\pm2\ldots$\h, and  $\beta=T^{-1}$.
The quantities $E_\pm$ and $f(z)$ are given by
\bea
E_\pm\h=\h\h\frac{1}{2m}
\left(\bk'\h\pm\h\h\frac{\bk}{2}\right)^2,\nn
\eea
\bea
f(z)\h=\h\h\frac{1}{1\h+\h e^{z-\mu/T}}\h.\nn
\eea

The induced Chern-Simons terms corresponding to the spin-up and spin-down
fermions are mutually canceled out in accordance with the parity invariance.

From the definition of $E_\pm$ it follows that $E_\pm>\h E_\mp$ for
sgn$(\bk\bk')=\pm1$, and since $f(z)$ is a decreasing function, we get
\bea
[\h f(\beta E_+)\h-\h f(\beta E_-)\h](\bk\bk')\h<\h0\h.
\eea
This implies that $G_{\tau\tau}>0$, and the corresponding contributions
to $W_2$ are positive. The rest part of $W_2$ is related with pure
magnetic configurations and determines the two-point function
\bea
\Lambda(x_1,x_2)\h=\h\h\frac{\d^2W_2}{\d B(x_1)\d B(x_2)}\h,\nn
\eea
which, up to a constant factor, is the susceptibility of
the perturbative ground state.

Because of the translational invariance, it is convenient to analyze
the eigenvalues of $\Lambda(x_1,x_2)$ in the Fourier representation.
These eigenvalues can be written in the following form
\bea
\lambda(\xi,\bk,T)\h=\h1\h+\h\h\frac{e^2}{m}
\h\h G(\xi,\bk,T)\h,
\eea
where $G(\xi,\bk,T)$ is given by the Eq. (1).

From these expressions one can detect the possibility of the existence of
negative eigenvalues leading to the magnetic instability. In fact, due to the
relation (2), the $\s$-dependent part of $G$ is negative, and for sufficiently
large values of $\s^2$ the quantity $G(\xi,\bk,T)$ will become negative at
least for some $\xi$ and $\bk$.

In order, to expose the underlying physics, consider the dense matter of
spinless fermions. In the absence of gauge fields fermions are organized in
a Fermi sphere. Turning on the homogeneous magnetic field, fermions will be
rearranged into the Landau levels, and the orbital diamagnetism will lead to
the increase of the energy of the system. Let us now attach the spin to these
fermions. Inclusion of the spin with positive (negative) $\sigma eB$ decreases
(increases) the energy. Due to the chemical equilibrium between the opposite
spin subsystems, the partial density of fermions with $\sigma eB>0$ exceeds
the one corresponding to $\sigma eB<0$. Therefore, the spin effects lead to
the spin paramagnetism promoting the decrease of the energy. If $\sigma^2$ is
large enough, then the paramagnetism dominates over the diamagnetism and causes
the overall decrease of the energy of the fermion system.

Further, for sufficiently large values of $e^2/m$, the negative values
of $G$ will dominate in Eq. (3), and for some $\xi$ and $\bk$ we shall
get $\lambda<0$.

In the light of these arguments we can formulate two main points required for
the existence of negative modes:
\begin{itemize}
\item[(i)]
the magnitude of $\s$ must be large enough to generate the instability
of the fermion matter.
\item[(ii)]
the value of the fraction $e^2/m$ must be large enough to guarantee that the
matter instability will not be compensated by the gauge field contributions.
\end{itemize}

In the following two subsections we present more detailed analysis of the
negative modes for static $(\xi=0)$ and nonstatic $(\xi\ne0)$ cases,
respectively.

\subsection{Static case}

Let us first explore the point (i) concerning the negative values of $G$.
As it is shown in the Appendix, the quantity $G(0,k,T)$ can be written
in the following form
\bea
G(0,k,T)\h=\h\h\int_0^1\frac{ds}{4\pi}\h\h
\frac{s^2\h-\h\s^2}{1\h+\h e^{\zeta(1-s^2)-\mu/T}}\h,
\eea
where $\zeta=k^2/8mT$ and $k=|\bk|$.

Assuming that the total particle density is fixed, we express the chemical
potential in terms of other parameters. The defining equation is
$\la\psi_\uparrow^*\psi_\uparrow^{}\ra
+\la\psi_\downarrow^*\psi_\downarrow^{}\ra=n_e$, with
\bea
\la\h\psi_\uparrow^*\psi_\uparrow^{}\h\ra\h=\h
\la\h\psi_\downarrow^*\psi_\downarrow^{}\h\ra\h=
\h\h\frac{mT}{2\pi}\h\h\ln\h(1\h+\h e^{\mu/T})\h,\nn
\eea
and consequently
\bea
\mu\h=\h T\h\ln\left(e^{1/\Theta}\h-\h1\right)\h,\nn
\eea
where $\Theta=mT/\pi n_e$.

Consider first the zero temperature limit. In that case the Fermi distribution
function becomes steplike, and we get
\bea
G(0,k,0)\h=\h\h
\frac{1\h-\h3\h\s^2}{12\pi}\h\h-\h\h\frac{1}{12\pi}
\sqrt{1\h-\h\h\frac{k_F^2}{k^2}}
\left[1\h-\h3\h\s^2\h-\h\h\frac{k_F^2}{k^2}\right]
\theta(k\h-\h k_F)\h,\nn
\eea
where the characteristic momentum $k_F$ is defined by
\bea
k^2_F\h=\h 8m\lim_{T\to0}\mu\h=\h 8\pi n_e\h.\nn
\eea

In Fig. 1(a) we present $G(0,k,0)$ for the different values of $\s^2$.
As one can be convinced, the negative eigenvalues appear only for $\s^2>1/3$.

\vspace*{0.2truecm}
\setlength{\unitlength}{1mm}
\begin{center}
\begin{picture}(175,70)
\put(16,0){\epsfxsize=140mm\epsfysize=70mm\epsfbox{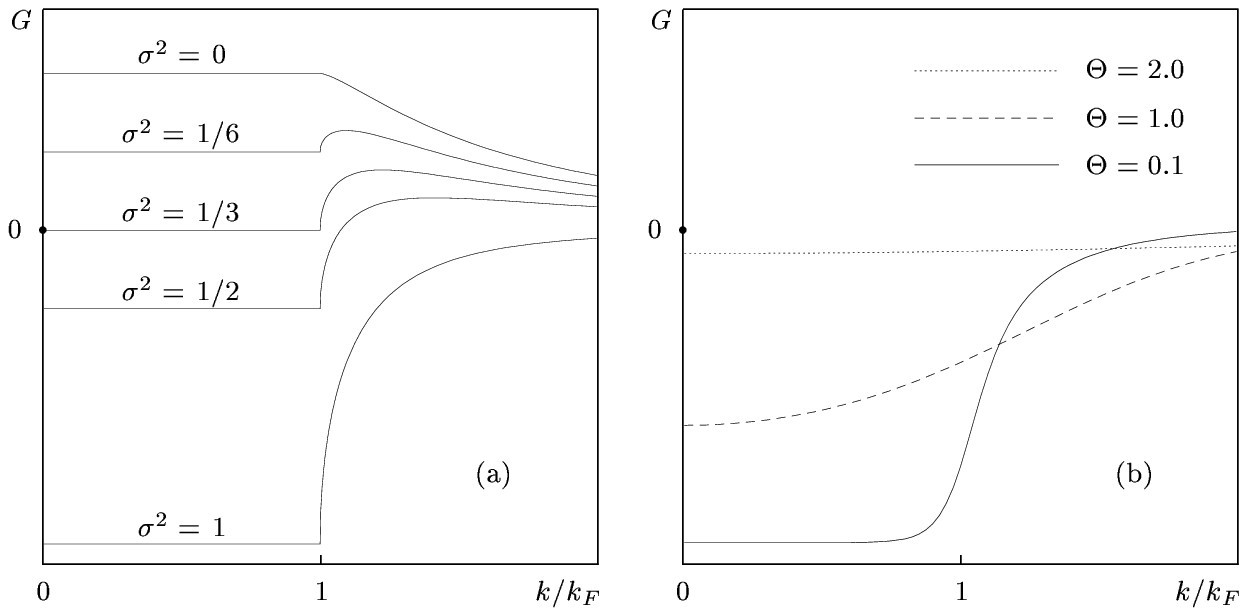}}
\end{picture}
\end{center}
\vspace*{-0.5truecm}
\begin{figure}
\caption{(a) $G(0,k,0)$ versus $k$;
         (b) $G(0,k,T)$ versus $k$, $\s^2=1$.}
\end{figure}

Consider now the point (ii) and trace out the condition guaranteeing
that the negative values of $G(0,k,0)$ will lead to $\lambda<0$.
For $\s^2>1/3$ the minimal value of $G(0,k,0)$ is given by
\bea
G(0,0<k<k_F,0)\h=\h-\h\h\frac{3\s^2-1}{12\pi}\h,\nn
\eea
and the required condition appears to be
\bea
\frac{m}{e^2}\h\h<\h\h\frac{3\s^2\h-\h1}{12\pi}\h.
\eea
In this case one obtains that $\lambda(0,k,0)<0$ for $0<k<k_c$, where $k_c$
is some critical value which is the solution to $\lambda(0,k,0)=0$. Remark,
that $k_c>k_F$.

The finite temperature behaviour of $G(0,k,T)$ is depicted in Fig. 1(b),
where the case of $\s^2=1$ is represented for different values of the
parameter $\Theta$. We see that $G(0,k,T)$ tends to zero when $T$ increases.
Therefore, the negative modes, observed at low temperatures, disappear in the
high temperature regime.

Remark, that the relation (5) can be realized only for $\s^2>1/3$ and
therefore, embraces the points (i) and (ii) simultaneously. Thus, the relation
(5) appears to be a sufficient condition for the existence of
$\lambda(0,k,T)<0$. Moreover, it is also a necessary one, since in the opposite
case $\lambda(0,k,T)$ is positive for all $k$ and $T$. In order to check up
this assertion, one can carry out some estimates. Note, that in (4) we have
$0<s<1$. In this interval $e^{\zeta(1-s^2)}>e^{\zeta(1-s)}$ and
$e^{\zeta(1-s^2)}<e^{\zeta(1-s^3)}$. Using these inequalities in (4),
one gets
\bea
\lambda(0,k,T)\h>\h1\h+\h\h\frac{e^2}{m}\h\h\frac{1\h-\h3\s^2}{12\pi}\h\h
\frac{1}{\zeta}\h\h\ln\frac{1\h+e^{\h\mu/T}}{1\h+e^{\h\mu/T\h-\h\zeta}}\h.\nn
\eea
Further, since $\zeta>0$, we have
\bea
0\h<\h\h\frac{1}{\zeta}\h\h\ln
\frac{1\h+e^{\h\mu/T}}{1\h+e^{\h\mu/T\h-\h\zeta}}\h\h<\h1\nn
\eea
and consequently,
\bea
\lambda(0,k,T)\h>\h1\h+\h\h\frac{e^2}{m}\h\h\frac{1\h-\h3\s^2}{12\pi}\h.\nn
\eea
This is a general relation valid for all values of $k$ and $T$, and
leading to $\lambda(0,k,T)>0$ when the condition (5) is not held.

\subsection{Nonstatic case}

Integrating out the polar angle in (1), the corresponding expression can be
presented in the following form
\bea
G(\xi,k,T)\h=\int_0^\infty\frac{dz}{4\pi}\left[\frac
{1-\s^2-z^2}{2}\h\h x\h+\h\h\frac{4m^2\xi^2z^2}{4m^2\xi^2+k^4(1+z)^2}\h\h
\frac{1}{x}\right](\varepsilon-1)f(\zeta z^2)\h,
\eea
\bea
\varepsilon\h=\h\sqrt{\h\frac{4m^2\xi^2+k^4(1+z)^2}{4m^2\xi^2+k^4(1-z)^2}}\h,
\hspace*{20mm}
x\h=\h\sqrt{\h\frac12\h\h\frac{4m^2\xi^2+k^4(1-z^2)}{4m^2\xi^2+k^4(1+z)^2}
\h\h+\h\h\frac{1}{2\h\varepsilon}}\h.\nn
\eea

In Fig. 2 we depict the behaviour of $G(\xi,k,T)$ for $\s^2=0$.
In that case paramagnetism is absent and the eigenvalue spectrum is
positively defined.

\vspace*{0.2truecm}
\setlength{\unitlength}{1mm}
\begin{center}
\begin{picture}(175,70)
\put(16,0){\epsfxsize=140mm\epsfysize=70mm\epsfbox{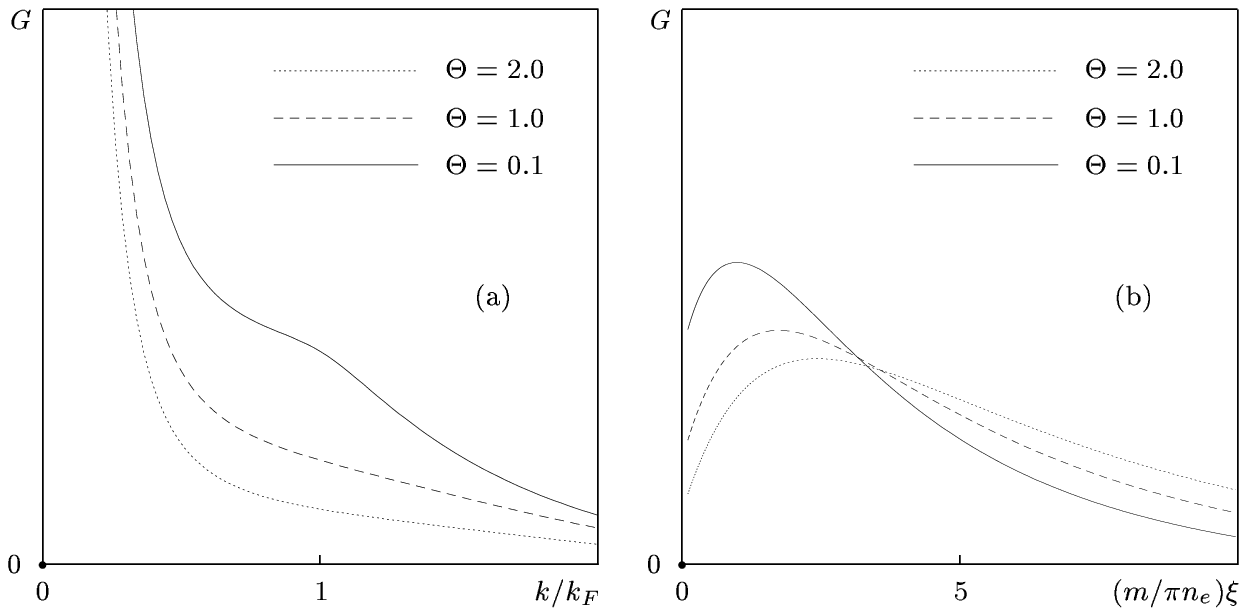}}
\end{picture}
\end{center}
\vspace*{-0.5truecm}
\begin{figure}
\caption{(a) $G(\xi,k,T)$ versus $k$, $\xi=0.1(\pi n_e/m)$, $\s^2=0$;
          (b) $G(\xi,k,T)$ versus $\xi$, $k=0.8k_F$, $\s^2=0$.}
\end{figure}

The case $\s^2=1$ is depicted in Fig. 3, where we observe that $G$
becomes negative for some values of $\xi$ and $k$.

\vspace*{0.2truecm}
\setlength{\unitlength}{1mm}
\begin{center}
\begin{picture}(175,70)
\put(16,0){\epsfxsize=140mm\epsfysize=70mm\epsfbox{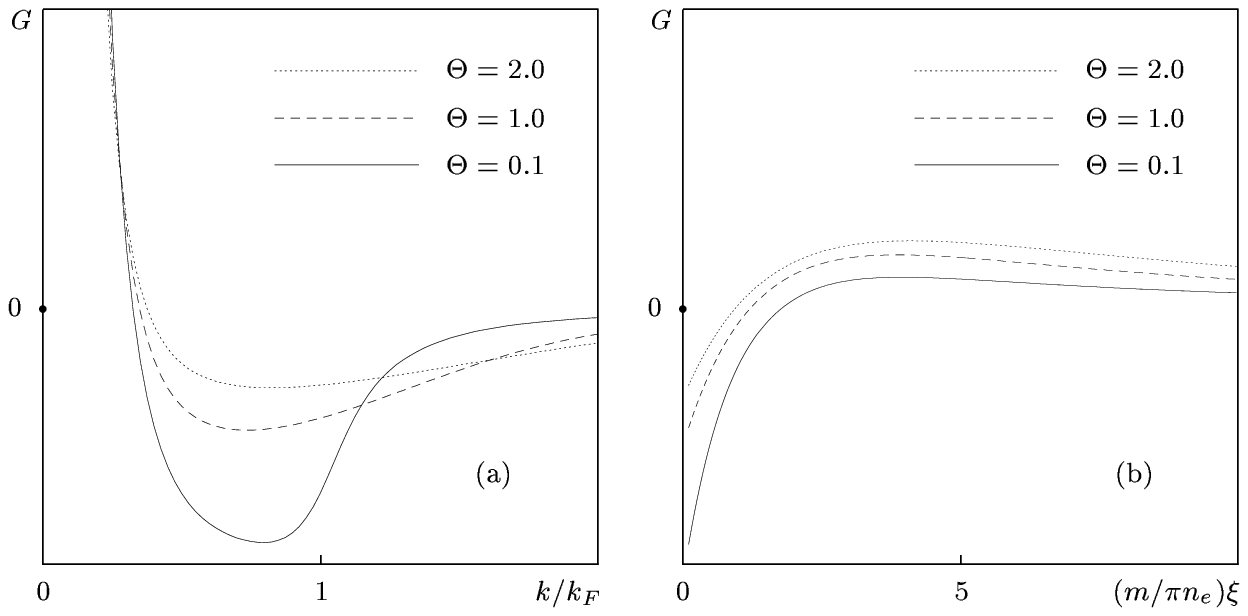}}
\end{picture}
\end{center}
\vspace*{-0.5truecm}
\begin{figure}
\caption{(a) $G(\xi,k,T)$ versus $k$, $\xi=0.1(\pi n_e/m)$, $\s^2=1$;
         (b) $G(\xi,k,T)$ versus $\xi$, $k=0.8k_F$, $\s^2=1$.}
\end{figure}

From the expression (6) one can extract the asymptotic properties
of $G(\xi,k,T)$ for $k\to\infty$ and $\xi\to\infty$. In the large-$k$
limit we obtain
\bea
G\h\rightarrow\h2\h\h\frac{1\h-\h\s^2}{2\pi\d}
\h\h\frac{mT}{k^2}\h\h\ln(1\h+\h e^{\h\mu/T})\5\5for\5\5\s^2\ne1\nn
\eea
\bea
G\h\to\h8\h\h\frac{\d\h-\h2}{\pi\d^2}\h\h
\frac{m^2T^2}{k^4}\int_0^\infty
\ln(1\h+\h e^{\h-z+\mu/T})\h dz\5\5for\5\5\s^2=1\nn
\eea
where $\d=1+(2m\xi/k^2)^2$. These expressions are valid irrespectively whether
the fraction $\xi/k^2$ vanishes, diverges or stays finite when $k\to\infty$.

Consider the limit $\xi\to\infty$. The case when $k$ tends to infinity together
with $\xi$ has been already discussed above. Therefore, we assume that $k$ is
finite and get
\bea
G\h\to\h\h\frac{1\h-\h\s^2}{4\pi}\h\h
\frac{Tk^2}{m\xi^2}\h\h\ln(1\h+\h e^{\h\mu/T})\h
+\h\h\frac{2}{\pi}\h\h\frac{T^2}{\xi^2}
\int_0^\infty\ln(1\h+\h e^{\h-z+\mu/T})\h dz\h.\nn
\eea

The asymptotic relations imply that the negative values of $\lambda$ can
be located in a finite region of the $(\xi,k)$-plane.

Figures 2 and 3 demonstrate that the magnitude of the negative values of
$G$ tend to zero as $T$ increases. Therefore, in the high temperature
regime we get $\lambda>0$.

\section{Acknowledgments}

We would like to thank G. Japaridze, L. O'Raifeartaigh, V. Rubakov
and P. Sodano for helpful discussions. The work was supported by the grant
Intas-Georgia 97-1340.

\section{Appendix}
\setcounter{equation}{0}
\def\theequation{A.\arabic{equation}}

Here we comment on the calculation of $G(0,\bk,T)$. The common details
are shown for its $\s$-dependent part:
\bea
G_{\tau\tau}(0,\bk,T)\h=\h-\int\frac{\dk'}{2\pi^2}
\h\h\frac{f(\beta E_+)-f(\beta E_-)}{\bk\bk'}\h.\nn
\eea
Due to the singularity, the $E_\pm$ parts are separately divergent and the
integral cannot be decoupled in the corresponding way.

Expanding $f(\beta E_{\pm})$ in powers of $\bk\bk'$, we use the polar variables
and integrating over $\bk'$ get
\bea
G_{\tau\tau}(0,k,T)\h=\h\h\frac{1}{\pi}\sum_{n=0}^\infty
\frac{(-1)^n\zeta^n}{n!\h(2n+1)}\h\h f^{\h(n)}(\zeta)\h,\nn
\eea
where $\zeta=k^2/8mT$. Expand $f^{\h(n)}(\zeta)$ and arrange the powers
of $\zeta$. Besides, we use $f^{\h(j)}(0)=(-1)^j\phi^{(j)}(\mu/T)$ with
$\phi(z)=(1+e^{-z})^{-1}$ and get
\bea
G_{\tau\tau}(0,k,T)\h=\h\h\frac{1}{2\sqrt{\pi}} \sum_{n=0}^\infty
\frac{(-1)^n\zeta^n}{\Gamma(n+3/2)}\h\h\phi^{(n)}(\mu/T)\h.
\eea

Analyze the structure of $\phi^{(n)}(z)$. From $\phi\h'(z)=\phi(1-\phi)$
it follows that $\phi^{(n)}(z)$ can be presented as an $(n+1)$'th order
polynomial with respect to $\phi$
\bea
\phi^{(n)}(z)\h=\h P_{n+1}[\phi(z)]\h.
\eea
These polynomials satisfy the recurrency relation
$P_{n+1}=\phi(1-\phi)(dP_n/d\phi)$ with $P_1[\phi]=\phi$.
By the direct calculation it can be checked up that the solution to
this recurrency chain appears as
\bea
P_{n+1}[\phi]\h=\sum_{k=0}^\infty
\sum_{l=0}^k(-1)^l(l+1)^nC^l_k\h\phi^{k+1}\h.
\eea

Substituting Eqs. (A.2) and (A.3) into Eq. (A.1), we use
\bea
\sum_{n=0}^\infty\frac{(-1)^n x^n}{\Gamma(n+\nu+1)}\h\h=
\frac{2}{\Gamma(\nu)}\int_0^1e^{-x(1-s^2)}s^{2\nu-1}ds\h,\nn
\eea
and summing up over $l$, get
\bea
G_{\tau\tau}(0,k,T)\h=\int_0^1
\frac{dt}{\pi}\h\h e^{-\zeta(1-s^2)}\sum_{k=0}^\infty
\frac{[1-e^{-\zeta(1-s^2)}]^k}{[1+e^{-\mu/T}]^{k+1}}\h.\nn
\eea

The infinite sum over $k$ converges, yielding
\bea
G_{\tau\tau}(0,k,T)\h=\h\h\frac{1}{\pi}
\int_0^1\frac{ds}{1+e^{\h\zeta(1-s^2)-\mu/T}}\h.
\eea

Performing the similar manipulations for the $\s$-independent part of $G$
and combining with (A.4), we get (4).


\begin{thebibliography}{000}
\bibitem{1}
P. Cea, {\it Phys. Rev.} {\bf D32}, 2785 (1985);
        {\it ibid.} {\bf D34}, 3229 (1986).

\bibitem{2}
Y. Hosotani, {\it Phys. Lett.} {\bf B319}, 332 (1993);
             {\it Phys. Rev.} {\bf D51}, 2022 (1995).

\bibitem{3}
D. Weselowski and Y. Hosotani, {\it Phys. Lett.} {\bf B354}, 396 (1995).

\bibitem{4}
V. Zeitlin, {\it Mod. Phys. Lett.} {\bf A12}, 877 (1997).

\bibitem{5}
S. Kanemura and T. Matsushita, {\it Phys. Rev.} {\bf D56}, 1021 (1997).

\bibitem{6}
P. Cea and L. Tedesco, {\it Phys. Lett.} {\bf B425}, 345 (1998).

\bibitem{7}
M. Eliashvili and G. Tsitsishvili, {\it Phys. Rev.} {\bf B57}, 2713 (1998).
\end{thebibliography}
\end{document}